\documentclass[aps,arxiv,preprint,superscriptaddress,groupedaddress]{revtex4-1}  
\usepackage{graphicx}
\usepackage{dcolumn}
\usepackage{bm}
\usepackage{amsmath}
\usepackage{amssymb}
\usepackage{mathtools}
\usepackage{float}
\usepackage{listings}
\usepackage[english]{babel}
\usepackage{graphicx}
\usepackage{xcolor}
\usepackage[version=4]{mhchem}

\newlength{\singlecol}
\newlength{\doublecol}
\newcommand{\red}[1]{\textcolor{red}{#1}}

\newcommand{\figref}[2][]{\hyperref[#2]{Figure~\ref{#2}#1}}
\newcommand{\tabref}[2][]{\hyperref[#2]{Table~\ref{#2}#1}}

\begin{document}
\title{%
\textsc{THIS MANUSCRIPT IS WORK IN PROGRESS}\\
Machine Learning Potential Powered Insights into the Mechanical Stability of Amorphous Li-Si Alloys}


\author{Zixiong Wei}%
\affiliation%
{Materials Chemistry and Catalysis, Debye Institute for Nanomaterials Science, Department of Chemistry, Utrecht University, Universiteitsweg 99, 3584 CG, Utrecht, Netherlands}
\author{Nongnuch Artrith}%
\email%
{n.artrith@uu.nl}
\affiliation{Materials Chemistry and Catalysis, Debye Institute for Nanomaterials Science, Department of Chemistry, Utrecht University, Universiteitsweg 99, 3584 CG, Utrecht, Netherlands}
\date{\today}

\begin{abstract}

%
Understanding the mechanical properties of solid-state materials on the atomic scale is crucial for the development of novel materials.
For instance, amorphous LiSi alloys, \ce{a-Li_xSi} are attractive anode materials for solid-state Li-ion batteries but suffer from mechanical instabilities due to significant volume variations when the Li content changes.
A fundamental understanding of the mechanical behavior of such systems is necessary to address their poor mechanical integrity.
Experimental characterization methods provide insufficient information to elaborate dynamic mechanical degradation mechanisms on the atomic scale, and first-principles methods that can describe atomic details, such as density functional theory (DFT), are computationally too demanding to access the system sizes necessary to model mechanical phenomena.
Machine learning potentials (MLPs) can overcome the computational constraints of traditional DFT-based simulations and enable large-scale simulations with high accuracy and efficiency.

Here, we provide a step-by-step tutorial on how to develop and apply MLPs to the investigation of mechanical properties in materials systems from bulk to nanoparticles in different compositions with Li-Si alloys as example.
By training on a large and comprehensive data set (around 45,000 DFT structures) with the atomic energy network (ænet) package accelerated by PyTorch, a general and robust MLP is constructed that can reproduce results consistent with previous experimental observations.
We demonstrate how to apply the MLP to realistic structures to visualize the deformation mechanism and determine the origin of mechanical instabilities caused by fracturing.
By providing detailed insights into the different steps required to simulate the mechanical behaviour of Li–Si systems, this work aims to establish MLP-based simulations as a tool to understand the mechanical behavior in different materials systems at the atomic scale.
\end{abstract}

\maketitle


\section{Introduction}

The flammable liquid electrolyte in conventional lithium-ion batteries is a safety risk, and solid-state batteries that instead use solid Li-ion conductors can increase the safety while also enabling enhanced current densities and faster charging times~\red{[ref]}.

It has recently been demonstrated that nanostructured silicon can replace graphite as anode material in solid-state batteries to substantially increase the energy density~\red{[ref]}.
The most significant challenge for the commercialization of silicon anodes remain their low mechanical stability as the material undergoes large volume expansions and contractions during lithiation and delithiation~\red{[ref]}.
Moreover, recent studies emphasized the critical role of mechanics in solid-state batteries as the stress and strain resulting from battery cycling is a common failure mode~\cite{science-battery-mechanics}~\red{[ref]}. 

Understanding dynamic failure mechanisms due to evolving mechanical quantities is crucial for the further development of solid-state batteries, but obtaining such insight from experimental characterization is extremely challenging.
Instead, atomic-scale simulations allow monitoring the evolution of stress and strain and are often better suited to elucidate the reasons for poor mechanical integrity caused by mechanical deformations on the atomic scale.

Molecular dynamics (MD) simulations with machine learning potentials (MLPs) provide a large-scale simulation approach with high accuracy and computational efficiency~\red{[refs]}.
MLPs trained on first principles density-functional theory (DFT) calculations can be essentially as accurate as the reference method while overcoming the intrinsic limitations in system sizes imposed by the great computational demand of DFT.
However, obtaining a reliable and robust MLP that accurately describes the mechanical properties of a given materials system is a complex process and care has to be taken to avoid method-intrinsic issues such as overfitting and extrapolation.
Here, we outline a step-by-step tutorial to guide through the construction and application of MLPs for the investigation of mechanical properties.



\section{Model and Methodology}

In this study, both amorphous bulk and nanoparticle \ce{a-Li_xSi} models were constructed to investigate the mechanical properties of the Li-Si system.
The MLPs were trained with the open-source Atomic Energy Network (ænet) software package~\cite{aenet1,aenet2,aenet3}.
The training process can be greatly accelerated by fully utilizing the Graphics Processing Unit's (GPU) computing power.
To achieve this, a GPU-supported extension of the ænet code was used as a substitute for the original training step~\cite{aenet-pytorch}.
All MD simulations were performed using the publicly available simulation code package Large-Scale Atomic/Molecular Massively Parallel Simulator (LAMMPS)~\cite{lammps-1,lammps-new}, with the OVITO (Open Visualization Tool) package used for visualization of the MD results~\cite{ovito}.


\subsection{Data set and MLP based on artificial neural networks}

\begin{figure*}[tbp]
  \centering
  \includegraphics[width=1.1\textwidth]{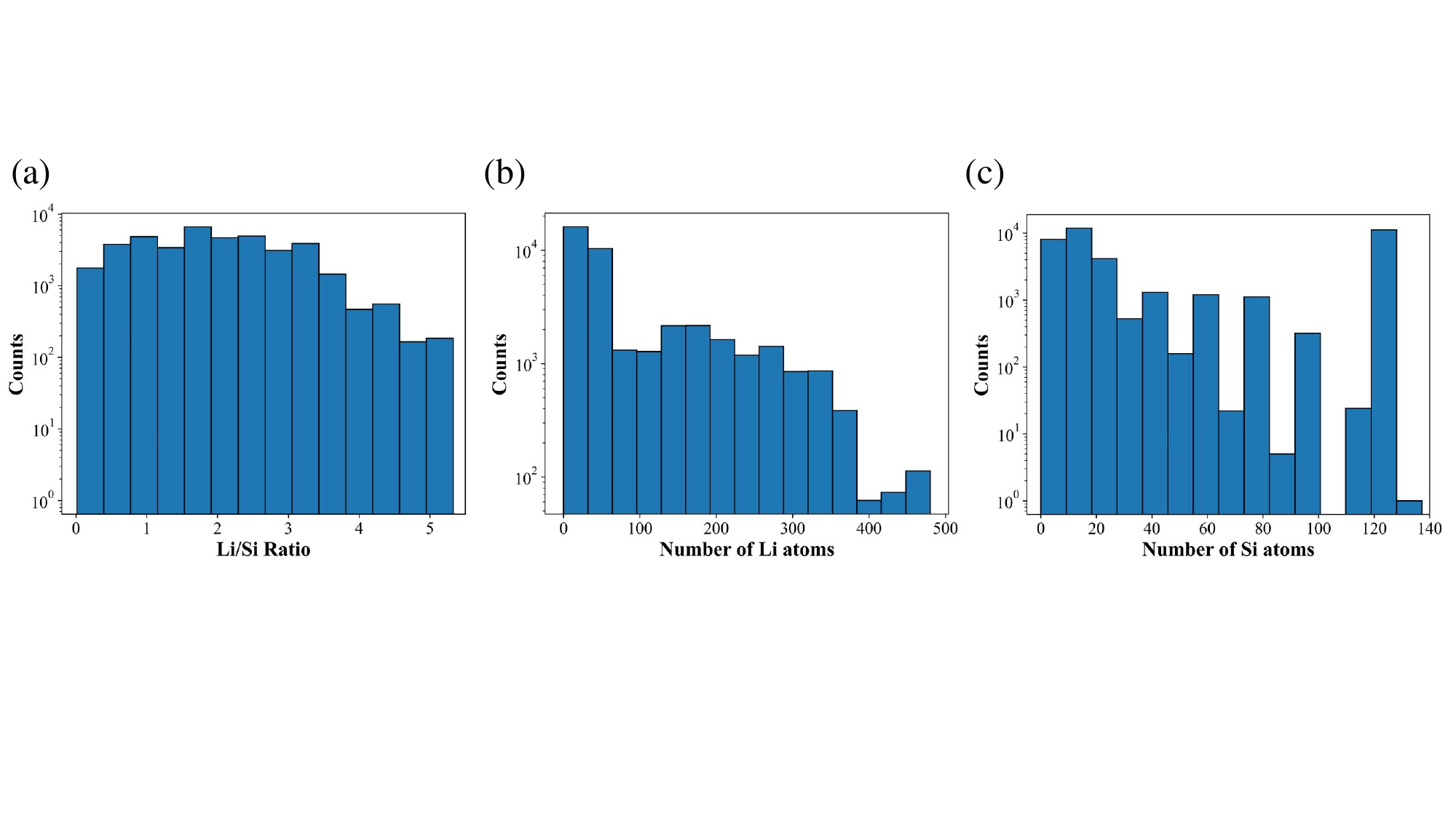}
  \vspace{-0.5\baselineskip}
  \caption{\label{fig: dataset}%
    \textbf{Data analysis of the 45,000 \ce{a-Li_xSi} reference structures.}
    \textbf{a}, Distribution of the Li/Si atom ratio across all structures of the MLP reference data set.
    \textbf{b}, Number of Li atoms in different structures.
    \textbf{c}, Number of Si atoms in different structures.
    }
\end{figure*}

For accurate and general MLPs, large, high-quality, and comprehensive training and testing data sets are required.
This study's data set was chosen from Artrith \textit{et al}~\cite{arxiv, phase-diagram}, which covers around 45,000 bulk, surface, and cluster Li-Si structures at various compositions.
An overview analysis of the data set is provided in \textbf{Fig.~\ref{fig: dataset}}, which shows the distribution of the Li/Si atom ratio and the number of Li/Si atoms in different structures.

Feed-forward artificial neural networks (ANN) were chosen as the model architecture for the MLPs.
Previous studies have shown that ANNs trained on this data set could provide a new approach for investigating the Li diffusivity in Si anodes, phase diagrams of \ce{a-Li_xSi}, and the delithiation process of LiSi nanoparticles\cite{arxiv,phase-diagram}.
The architecture of the feed-forward ANN in this work consisted of two hidden layers with 25 neurons each.
The hyperbolic tangent function (Tanh) was chosen as the activation function.
The local atomic environment and chemical species are represented by a descriptor constructed with the radial and angular distribution functions (RDF and ADF) whose coefficients are expanded by a basis set of Chebyshev polynomials~\cite{aenet2} that have the following form:
\begin{equation}
\begin{aligned}
& \operatorname{RDF}_i(r)=\sum_{\mathbf{R}_j \in \sigma_i^{R_{\mathrm{c}}}} \delta\left(r-R_{i j}\right) f_{\mathrm{c}}\left(R_{i j}\right) w_{t_j}, \\
& \operatorname{ADF}_i(\theta)=\sum_{\mathbf{R}_j, \mathbf{R}_k \in \sigma_i^{R_{\mathrm{c}}}} \delta\left(\theta-\theta_{i j k}\right) f_{\mathrm{c}}\left(R_{i j}\right) f_{\mathrm{c}}\left(R_{i k}\right) w_{t_j} w_{t_k},
\end{aligned}
\end{equation}
where $R_{i j}$ is the radial distribution between atoms $i$ and $j$; $\theta_{i j k}$ is the angular distribution for atoms $i$, $j$, and $k$; $w_{t_i}$ are the weights for chemical species $t_i$; and $f_{\mathrm{c}}\left(r\right)$ is a cutoff function:
\begin{equation}
f_{\mathrm{c}}(r)= \begin{cases}\frac{1}{2}\left[\cos \left(r \cdot \frac{\pi}{R_{\mathrm{c}}}\right)+1\right] & \text { for } r \leq R_c \\ 0, & \text { else }\end{cases}
\end{equation}

\subsection{MLP training and validation}

\begin{figure*}[tbp]
  \centering
  \includegraphics[width=1.1\textwidth]{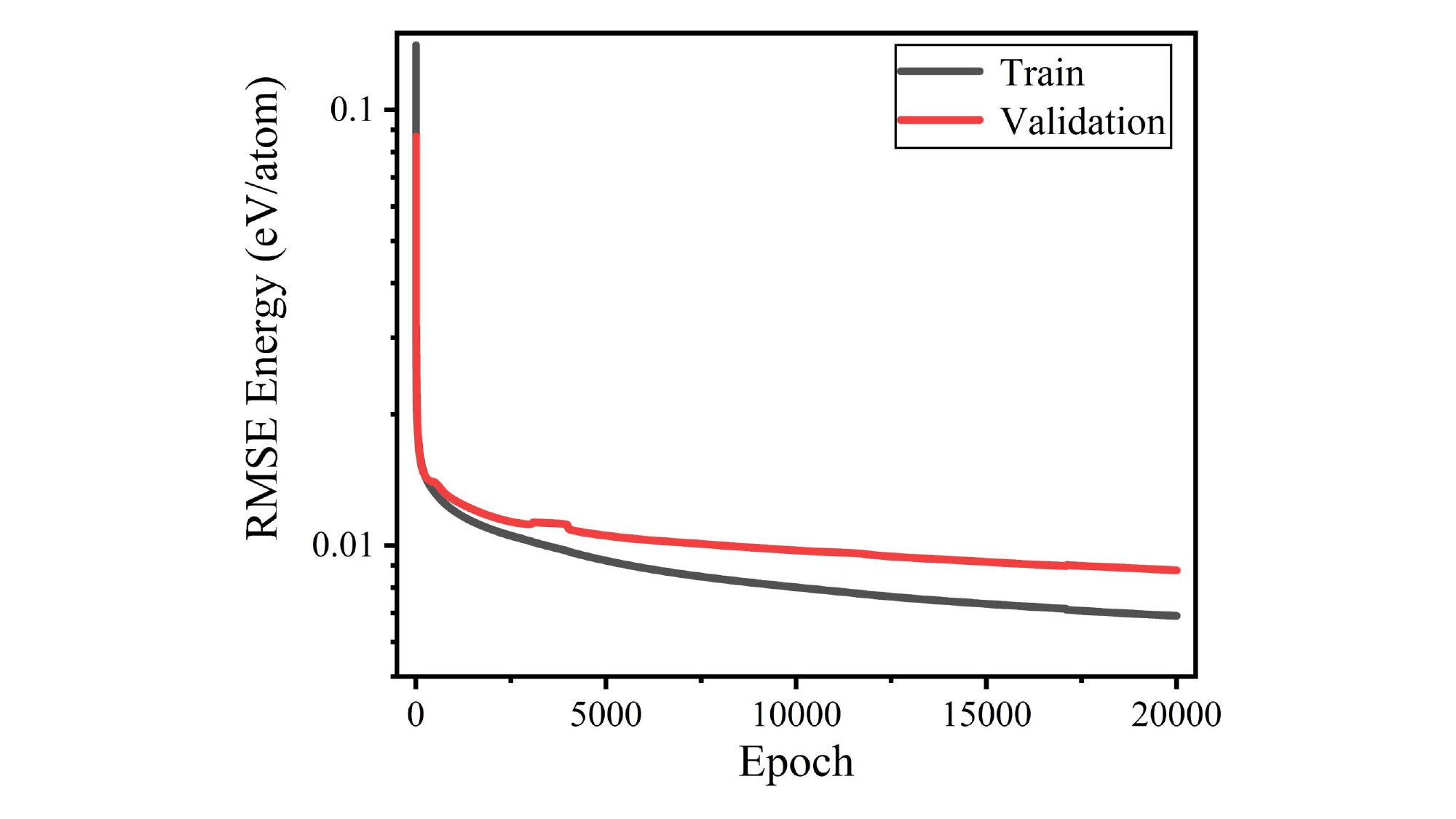}
  \vspace{-0.5\baselineskip}
  \caption{\label{fig: training}%
    \textbf{Evolution of root mean squared errors (RMSEs) during the training process.}
    }
\end{figure*}

The objective of training an MLP is to find suitable parameters in the nodes connecting the descriptor vector with the atomic energy output so that the potential can accurately interpolate the potential energy surface.
The hyperparameters were chosen as follows: 128 for the batch size, 0.0001 for the learning rate, and 10\% for the test set percentage.
The training process and validation result are shown in \textbf{Fig.~\ref{fig: training}}.

\begin{lstlisting}[language=bash]
TESTPERCENT 10
BATCHSIZE   128
METHOD
method=adamw   lr=0.0001
NETWORKS
! atom   network         hidden
! types  file-name       layers  nodes:activation
  Li     Li.pytorch.nn    2    25:tanh    25:tanh
  Si     Si.pytorch.nn    2    25:tanh    25:tanh
\end{lstlisting}


\subsection{Construction of Structure Models for Mechanical Simulations}

\begin{figure*}[tbp]
  \centering
  \includegraphics[width=\textwidth]{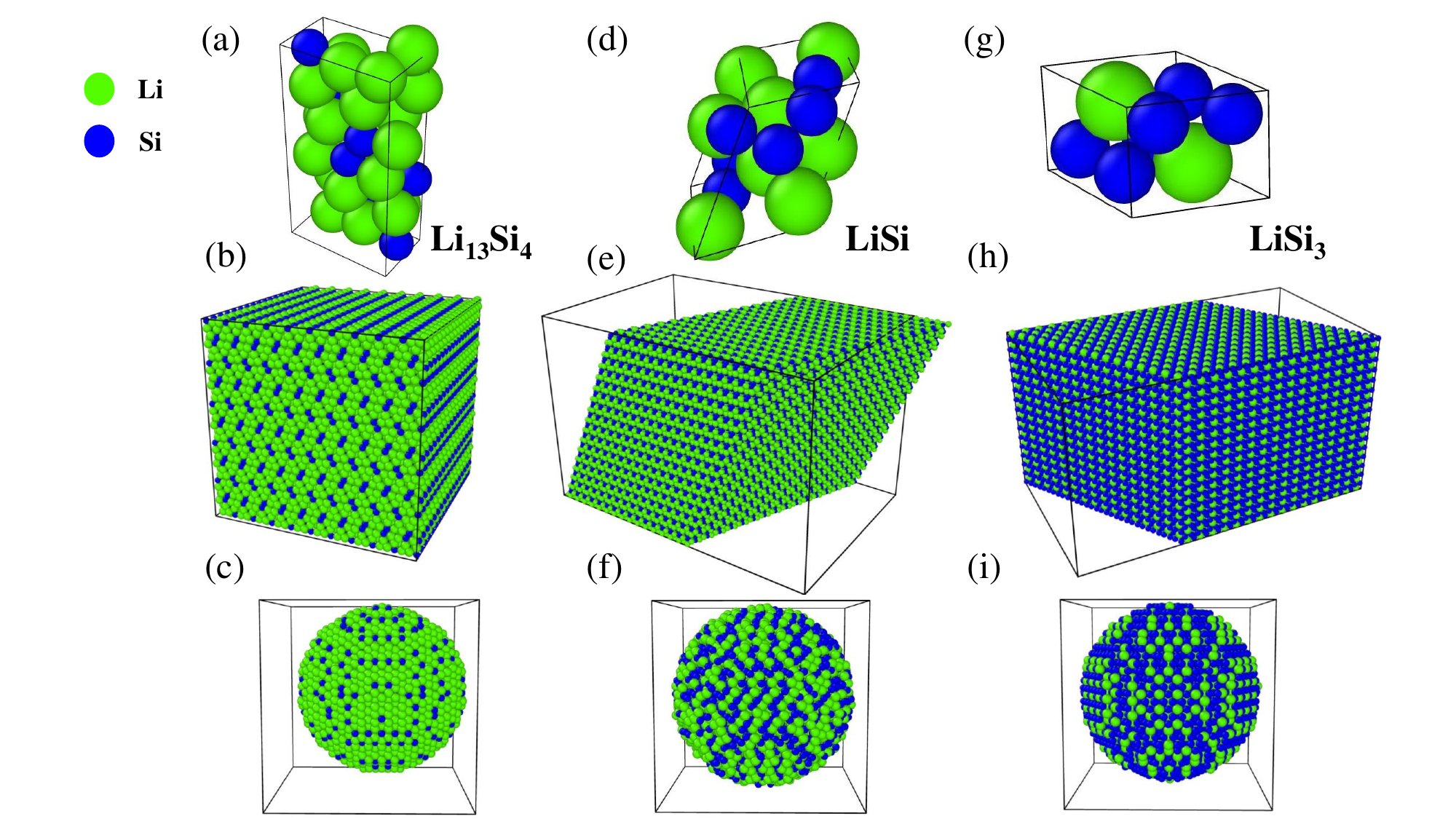}
  \vspace{-0.5\baselineskip}
  \caption{\label{fig: model1}%
    \textbf{Bulk and nanoparticle models constructed from different Li/Si compositions: \ce{Li13Si4}, LiSi, and \ce{LiSi3}.}
    \textbf{a}, \textbf{d}, \textbf{g}, Unit cells with different compositions.
    \textbf{b}, \textbf{e}, \textbf{h}, Expanded bulk structures as supercell with 28900 atoms, 65536 atoms and 32768 atoms.
    \textbf{c}, \textbf{f}, \textbf{i}, Sphere nanoparticles cut from bulk structures with 7344 atoms, 6625 atoms and 8397 atoms.
    }
\end{figure*}

The \ce{a-Li_xSi} alloy unit cells were downloaded from The Materials Project~\cite{material-project}.
To include different compositions, unit cells with the chemical formulas of \ce{Li_13Si_4}, \ce{LiSi}, and \ce{LiSi_3} were chosen.
As shown in \textbf{Fig.~\ref{fig: model1}}, they were expanded in three dimensions first, and then nanoparticles with a radius of 30 \(\text{\r{A}}\) were cut from bulk structures.

\begin{figure*}[tbp]
  \centering
  \includegraphics[width=\textwidth]{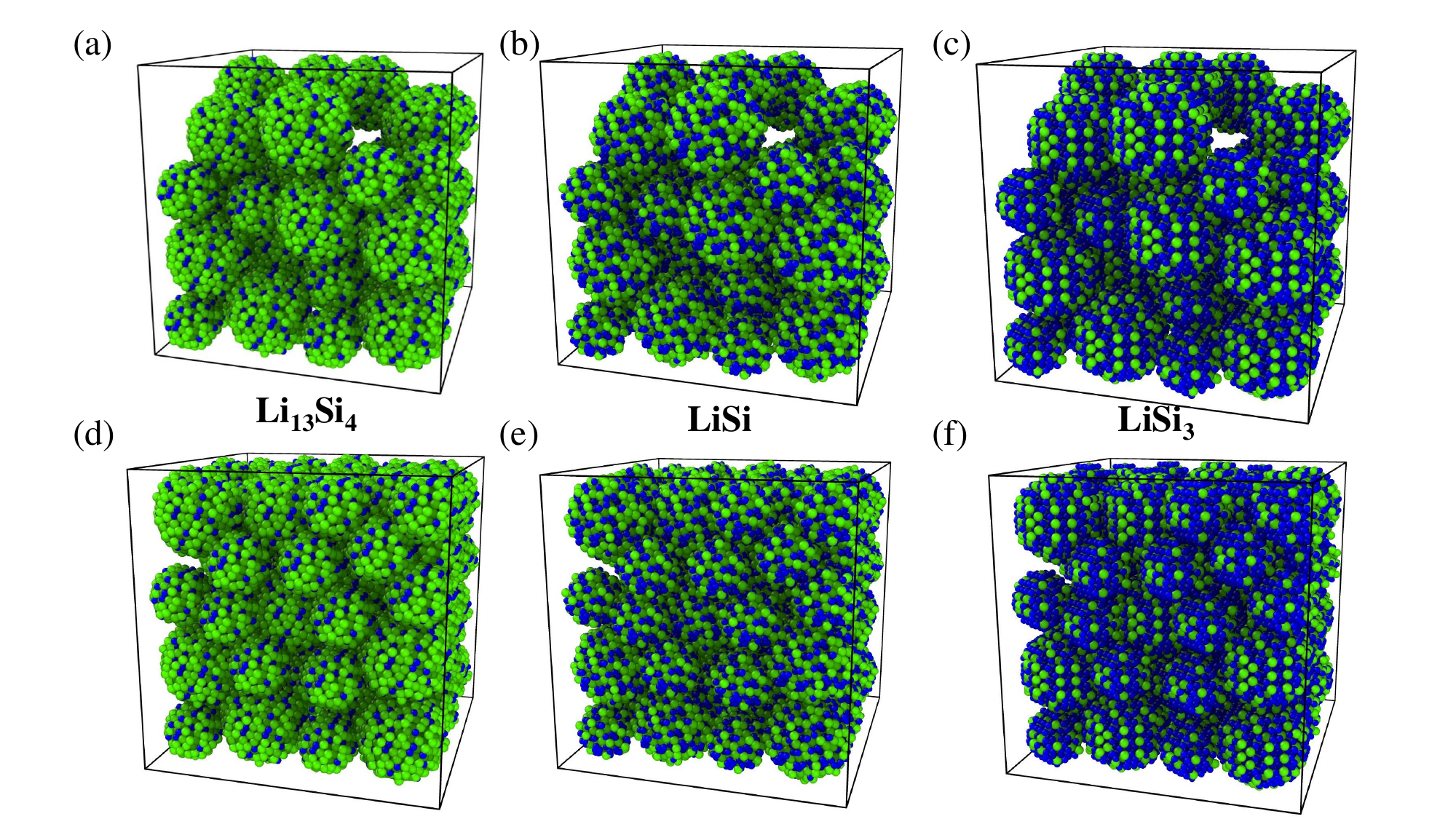}
  \vspace{-0.5\baselineskip}
  \caption{\label{fig: model2}%
    \textbf{Stacking models with different numbers of nanoparticles and compositions.}
    \textbf{a}, \textbf{b}, \textbf{c}, 50 particles in total with compositions of \ce{Li_13Si_4}, \ce{LiSi}, and \ce{LiSi_3} from left to right.
    \textbf{d}, \textbf{e}, \textbf{f}, 80 particles in total with compositions of \ce{Li_13Si_4}, \ce{LiSi}, and \ce{LiSi_3} from left to right.
    }
\end{figure*}

For the stacking models, inspired by the works of Srivastava \textit{et al}~\cite{model-particle} and Li \textit{et al}~\cite{model-replace}, a coarse-grained (CG) model was first constructed by pouring a certain number of granular particles with different sizes into a box with the size of 100 \(\text{\r{A}}\) $\times$ 100 \(\text{\r{A}}\) $\times$ 100 \(\text{\r{A}}\) using LAMMPS, and then all the particles were replaced by all-atom nanoparticles.
Two models containing 50 and 80 particles, respectively, were constructed, each of which was composed of particles with two sizes (R = 10 and 15 \(\text{\r{A}}\)).
A single nanoparticle with the composition Li13Si4, LiSi, and LiSi3 contains 7344, 6625, and 8397 total number of atoms, respectively.
As can be seen in \textbf{Fig.~\ref{fig: model2}}, there were 6 models in total with compositions shown in the middle of the figure.


\subsection{Molecular Dynamics Simulations with MLPs}

The energy of the systems was minimized with the conjugate gradient algorithm. All configurations were first relaxed in the canonical (\emph{NVT}) ensemble, then stretched in one dimension by a tensioning process in the isobaric-isothermal (\emph{NPT}) ensemble.
Periodic boundary conditions were applied in all directions.
A simulation time step of 1 fs was used. Simulations were run for 70 ps (20 ps relaxation and 50 ps stretching) in total at a temperature of 300 K at a strain rate of 1.0e10 (1/s).


\begin{lstlisting}[language=bash]
variable		dt equal 0.001
timestep		${dt}
#========== Force Field ==========
pair_style		aenet Li.pytorch.nn Si.pytorch.nn
pair_coeff		* *
#========== Equalibration ========
variable		Tdamp equal "v_dt*100"
fix		        1 all nvt temp 300.0 300.0 ${Tdamp}
#========== Deformation ==========
variable		Pdamp equal "v_dt*1000"
fix		        2 all npt temp 300.0 300.0 ${Tdamp} x 1 1 ${Pdamp} z 1 1 ${Pdamp} drag 1
\end{lstlisting}

\subsection {Running LAMMPS simulations using \ae{}net potentials}

The \ae{}net library files, as well as any other dependencies, need to be properly loaded (i.e. by setting the \$LD$\_$LIBRARY$\_$PATH). The LAMMPS input script also needs to be configured so as to use the \ae{}net \emph{pair style} and to specify which neural network parameter (\emph{.ann}) files to use.

A partial input example for water is provided below:

\begin{lstlisting}[language=bash]
units metal
mass 1   6.941
mass 2  28.0855
pair_style aenet Li.ann Si.ann
pair_coeff * *
\end{lstlisting}

For tensile stretch simulation of bulk structures, the model with the composition \ce{Li_13Si_4} was chosen, which consisted of around 100,000 atoms with the dimensions of 106 \(\text{\r{A}}\) $\times$ 126 \(\text{\r{A}}\) $\times$ 120 \(\text{\r{A}}\).
The simulation box was stretched at a constant engineering strain rate in the y direction, with the x and z directions able to deform to account for the Poisson effect. 

For simulations of nanoparticles and nanoparticle-stacked models, all three compositions were used.
All structures were divided into two parts of equal size, where one part was fixed and for the other one forces were applied in the direction in which they were stretched.

The virial stress tensor of an atom can be computed as~\cite{subramaniyan2008continuum}:
\begin{equation}
  \sigma=\frac{1}{\Omega}-\sum_\alpha m_\alpha \boldsymbol{v}_\alpha \otimes \boldsymbol{v}_\alpha+\frac{1}{2} \sum_{\alpha, \beta \neq \alpha} \boldsymbol{r}_{\alpha \beta} \otimes \boldsymbol{f}_{\alpha \beta} 
\end{equation}
where $\Omega$ is the volume of the selected observation space; $m_\alpha$ and $\boldsymbol{v}_\alpha$ are the mass and velocity of atom $\alpha$, respectively; $\boldsymbol{r}_{\alpha \beta}$ is the position vector from
atom $\alpha$ to atom $\beta$; and $\boldsymbol{f}_{\alpha \beta}$ is the force of atom $\beta$ on atom $\alpha$ in the conservative system.
Virial stress has been shown to be equivalent to continuum Cauchy stress.
Cauchy stress can be obtained by statistically averaging the virial stress of a specific region~\cite{lin2021debonding}.


\section{Results and Discussion}


\subsection{Deformation mechanism and mechanical properties of \ce{a-Li_13Si_4} bulk structure}

\begin{figure*}[tbp]
  \centering
  \includegraphics[width=\textwidth]{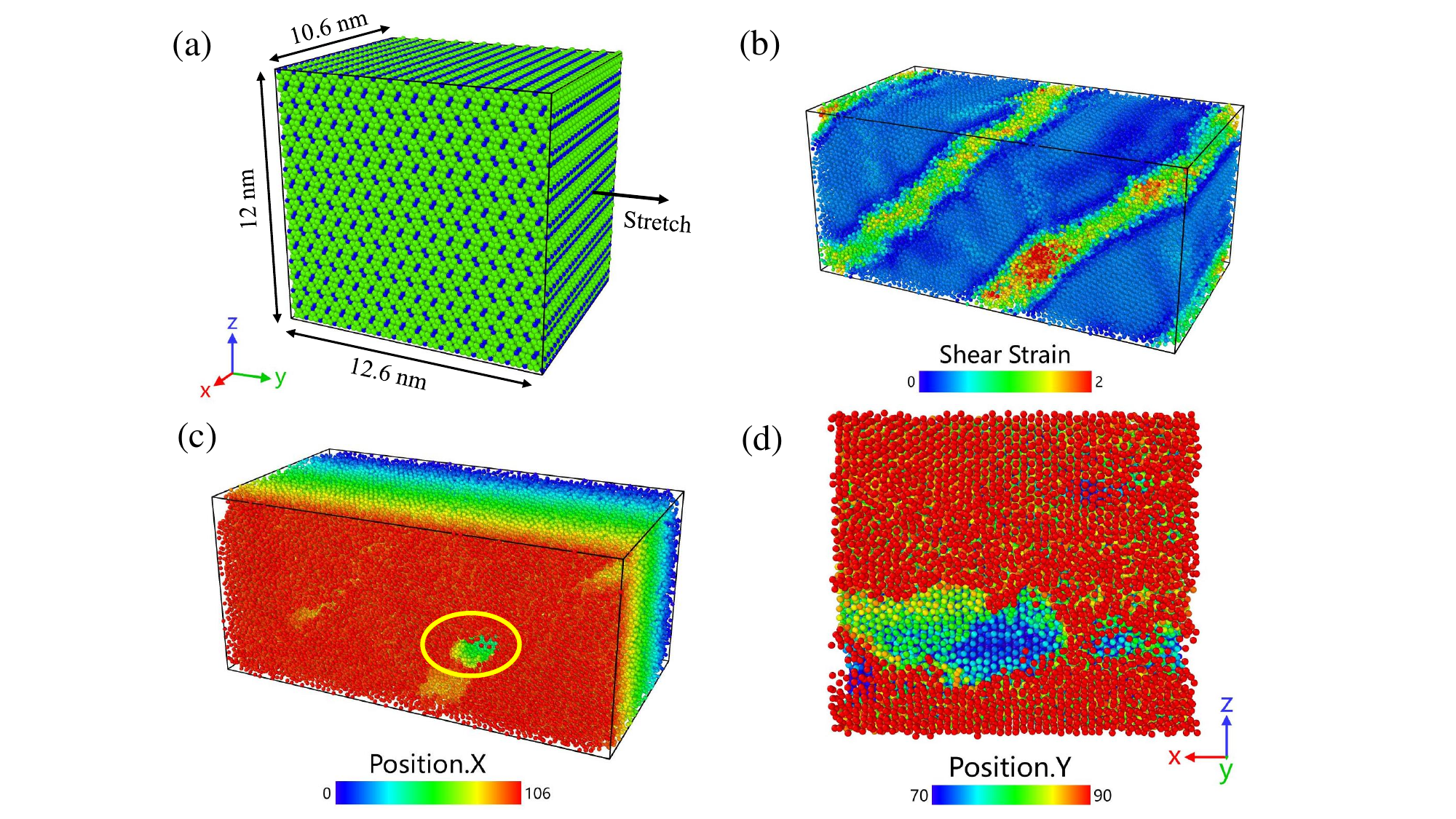}
  \vspace{-0.5\baselineskip}
  \caption{\label{fig: bulk}%
    \textbf{Deformation and fracturing of \ce{Li_13Si_4} bulk during stretching or compressing processes.}
    \textbf{a}, Initial MD models with T = 300 K, P = 1 bar.
    \textbf{b}, Shear strain distribution after stretching (1 fs/MD step and 70 ps simulation time).
    \textbf{c}, Formation of internal hole due to structure fracturing.
    \textbf{d}, Visualization of the hole's cross-section (circled by the yellow ellipse in \textbf{c}). \textbf{\textit{The units in c and d are in Angstrom}} }
\end{figure*}

As can be seen in \textbf{Fig.~\ref{fig: bulk}b}, the shear strain distribution in the final configuration (after a simulation time of 70 ps) of the bulk structure indicates that it deforms by forming a shear band.
Corresponding to the large shear strain within the shear band, the region highlighted by a yellow ellipse in \textbf{Fig.~\ref{fig: bulk}c} shows a hole caused by fracturing in the stretching process.
The cross-section along the xz plane is shown in \textbf{Fig.~\ref{fig: bulk}d}, where different colors represent atoms with different y coordinates.

\textbf{\textit{Comparing the stretching with compression, the bulk structure also shows the deformation mechanism that shear bands are formed. The strain in some regions is greater than that in other regions in the end, which means they are the first area to be fractured. As for the stress-strain curve, it has the similar shape as that in tensile stretching situation.}}

From the comparison of atoms with different colors, the sequence of positions of bond breakage could be estimated.
The bonds around the middle position in the cross-section break first, forming a small internal hole inside the bulk structure.
With the stretching process evolving, the crack tip gradually propagates to the edge surfaces of the bulk structure causing the hole to grow until finally penetrating through the bulk structure.

\begin{figure*}[tbp]
  \centering
  \includegraphics[width=1.3\textwidth]{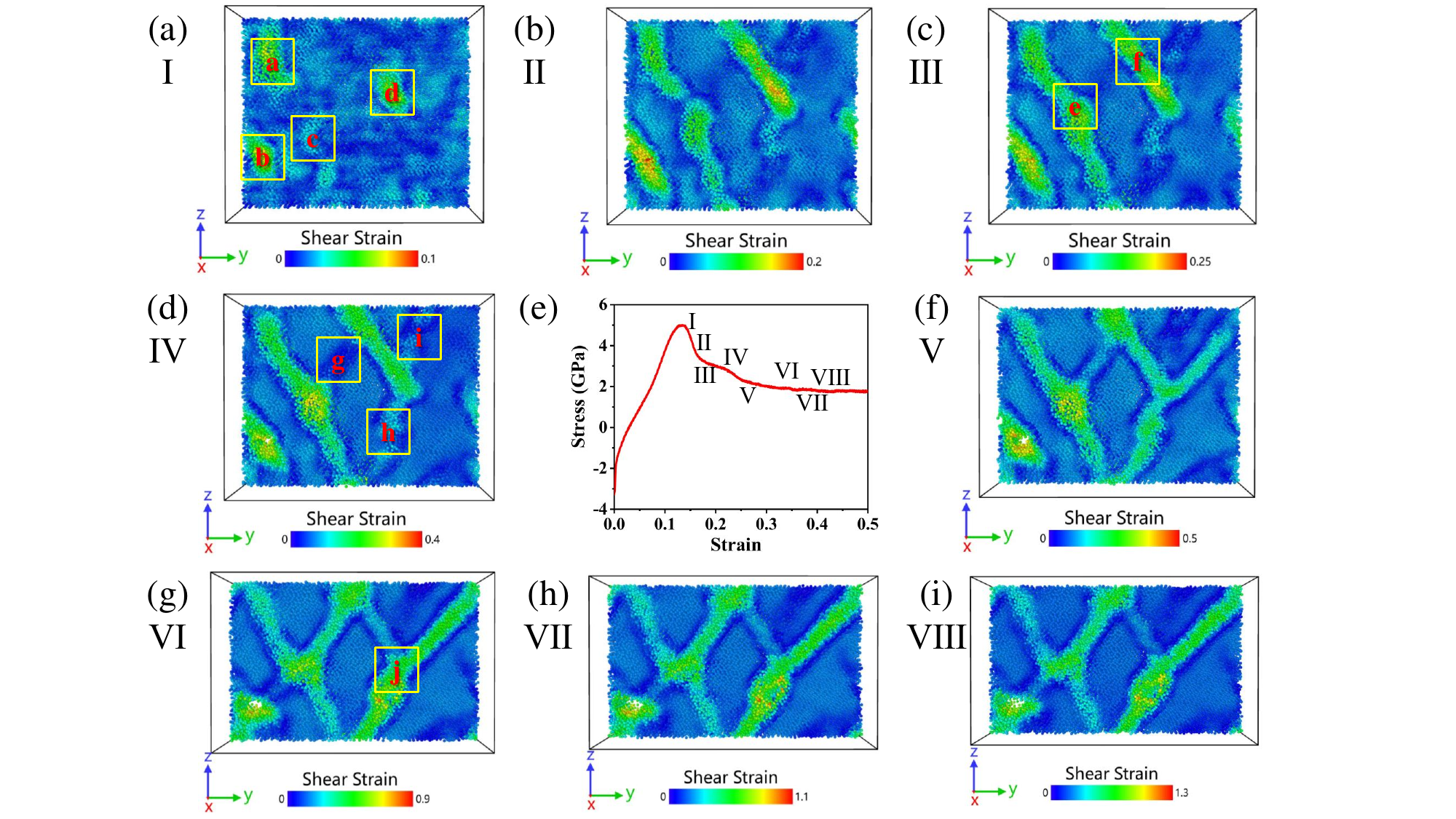}
  \vspace{-0.5\baselineskip}
  \caption{\label{fig: bulk-stress}%
    \textbf{Stress-strain curve from an MD simulation of the stretching process and snapshots of different states at different strains.}
    \textbf{a}, \textbf{b}, \textbf{c}, \textbf{d}, \textbf{f}, \textbf{g}, \textbf{h}, \textbf{i}, Cross-sections along the yz plane at different strains along the MD trajectory that correspond to the different points in \textbf{e}.
    }
\end{figure*}

The stress-strain curve of the bulk structure along with different shear strain distributions at different strains is shown in \textbf{Fig.~\ref{fig: bulk-stress}}.
The \ce{a-Li_13Si_4} bulk structure shows a typical elastic-plastic curve yet without an obvious plastic yielding stage.
It is worth noting that Young's modulus of our bulk structure is calculated as 45.04 GPa, which is quite close to the experimental value of 45.7 GPa measured by nanoindentation tests in reference~\cite{2017deformation}.

Recently, it was reported that amorphous shear bands found in crystalline materials act as drivers of plasticity~\cite{2023amorphous}.
Similarly, amorphous shear bands are also found in \ce{a-Li_13Si_4} bulk structures and the large plastic strain corresponds to the ductile behaviour.
Moreover, to visualize the distribution of shear strains at different strains, the cross-sections perpendicular to the x-axis are shown, which makes it possible to clarify the evolution of strain distribution and thus elucidate the deformation mechanism.

As shown in \textbf{Fig.~\ref{fig: bulk-stress}a}, four regions with strains larger than the rest of the other areas in the cross-section are denoted as a, b, c, and d.
With the bulk structure being further stretched, the shear strains in all regions tend to form shear bands and spread in directions at angles that are nearly parallel to each other (\textbf{Fig.~\ref{fig: bulk-stress}b}).
In \textbf{Fig.~\ref{fig: bulk-stress}c}, it can be seen that regions a and c gradually merge to form a long shear band denoted by e, where the meeting area is wider than the rest part.
Region d is expanded to become longer denoted by f.
Then, the strains in regions g, h, and i become larger (\textbf{Fig.~\ref{fig: bulk-stress}f}) and regions h and i also merge to become one band denoted by j that intersects with region f.
From \textbf{Fig.~\ref{fig: bulk-stress}g} to \textbf{Fig.~\ref{fig: bulk-stress}i}, it is shown that in the final stage, the strains in regions g and j keep becoming larger and finally exceed those in regions e and f.


\subsection{Deformation mechanism and mechanical properties of \ce{a-Li_xSi} nanoparticles}

\begin{figure*}[tbp]
  \centering
  \includegraphics[width=1.2\textwidth]{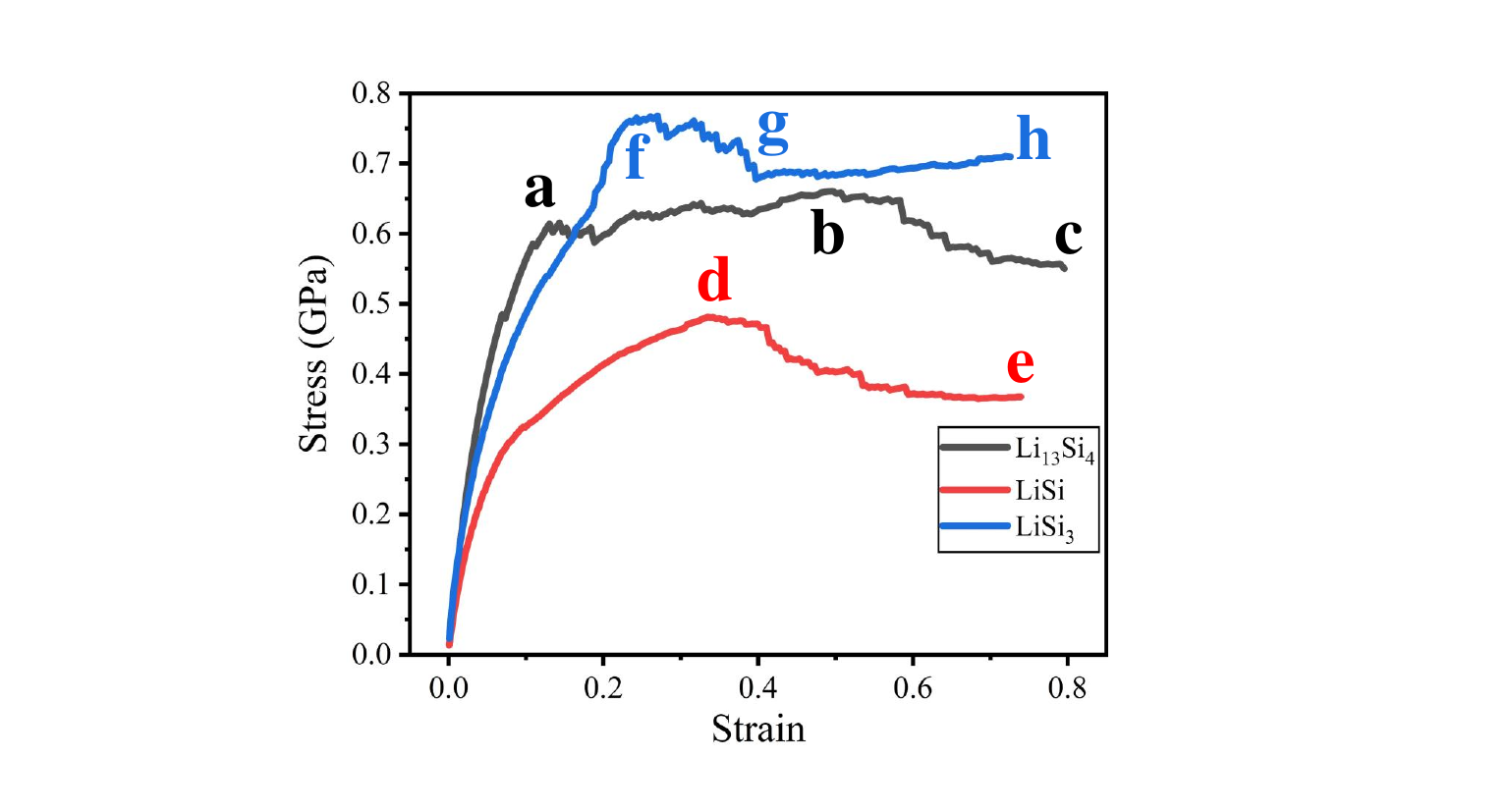}
  \vspace{-0.5\baselineskip}
  \caption{\label{fig: np-stress}%
    \textbf{Stress-strain curve of nanoparticles with different compositions in the stretching process.} The different regions along the curve are labeled according to the definition in Fig.~\ref{fig: nanoparticle}.
    }
\end{figure*}

The stress-strain curve of same-sized nanoparticles with different compositions is shown in \textbf{Fig.~\ref{fig: np-stress}}.
All curves show the typical elastic-plastic behaviours but with many noticeable differences which will be discussed in the following.
For \ce{a-Li_13Si_4} nanoparticle, the curve shows a short yet obvious plastic yielding stage around the area denoted a, followed by an obvious strain hardening stage before the area denoted b.
However, for both \ce{a-LiSi}, and \ce{a-LiSi_3} nanoparticles, no plastic yielding stages are found.
When it comes to the mechanics parameters, the ultimate tensile strength of \ce{a-LiSi_3} nanoparticle is the highest, larger than that of \ce{a-Li_13Si_4} nanoparticle, with \ce{a-LiSi} being the least.
For the stiffness, however, \ce{a-Li_13Si_4} nanoparticle has the largest value, \ce{a-LiSi_3} nanoparticle is the second, and \ce{a-LiSi} nanoparticle is the least one of all three.
It is worth noting that \ce{a-Li_13Si_4} nanoparticle has the best ductility and toughness as well.

\begin{figure*}[tbp]
  \centering
  \includegraphics[width=1.15\textwidth]{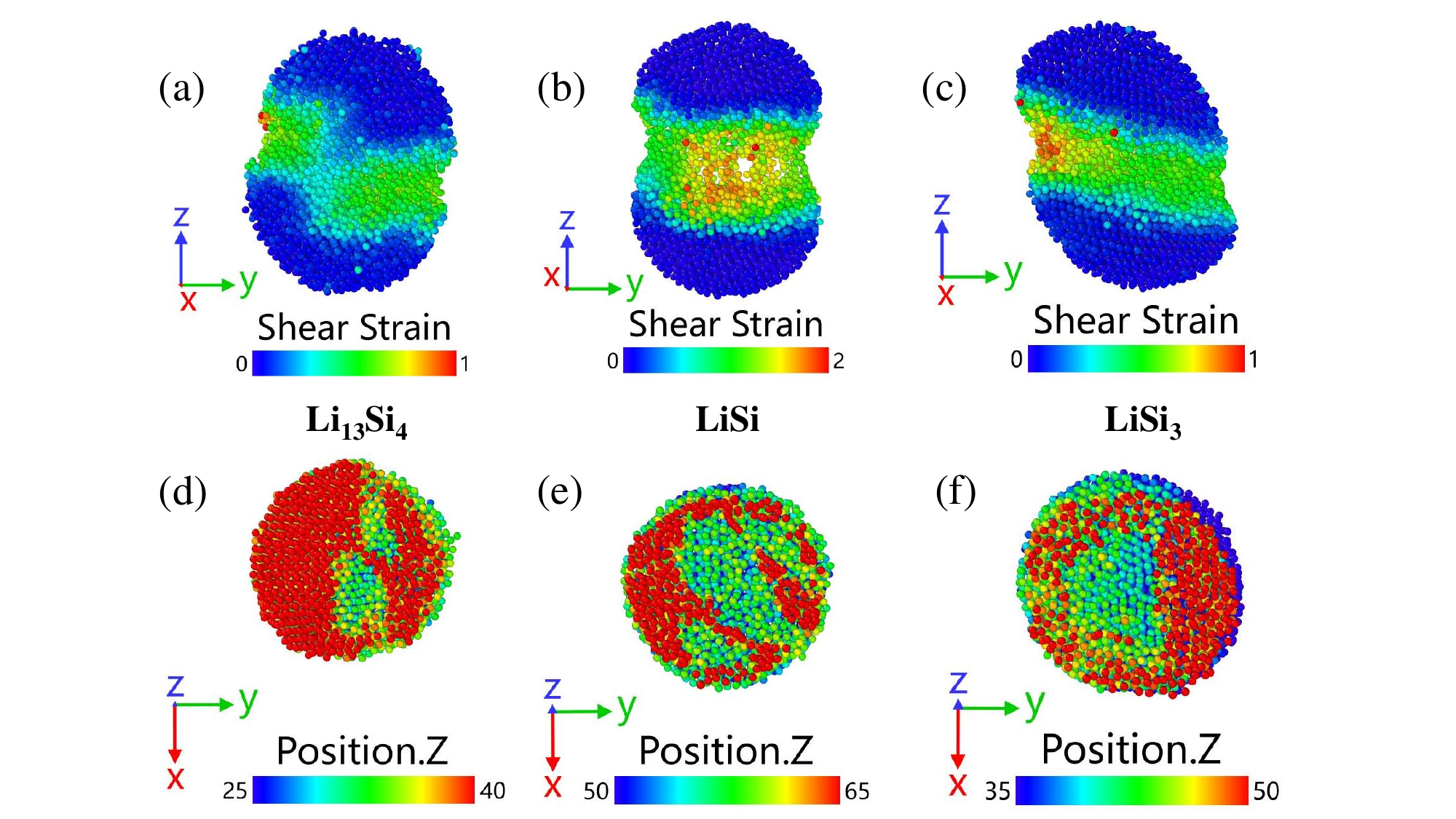}
  \vspace{-0.5\baselineskip}
  \caption{\label{fig: nanoparticle}%
    \textbf{Morphology of different nanoparticles in the stretching process.}
    \textbf{a}, \textbf{b}, \textbf{c}, Distribution of shear strains in different nanoparticles.
    \textbf{d}, \textbf{e}, \textbf{f}, Visualization of cross-sections of different nanoparticles.
    }
\end{figure*}

Relating different points in the stress-strain curves to their actual mechanical deformations provides an intuitive understanding of their behaviours subjected to the tension process.
In \textbf{Fig.~\ref{fig: nanoparticle}}, the distribution of shear strains and morphology of corresponding cross-sections are shown.
\textbf{Fig.~\ref{fig: nanoparticle}a} and \textbf{Fig.~\ref{fig: nanoparticle}d} correspond to the curve between a and b in \textbf{Fig.~\ref{fig: np-stress}}.
During a and b, the nanoparticle shows the formation of shear bands, which is the reason why the plastic strain is generated.
\textbf{Fig.~\ref{fig: nanoparticle}b} and \textbf{Fig.~\ref{fig: nanoparticle}c} are the states around f and d in
\textbf{Fig.~\ref{fig: np-stress}}, which not only shows the formation of shear bands but also shows the necking behaviours that are quite typical in materials.
It is interesting to point out that for the \ce{a-LiSi_3} nanoparticle it changes from sphere to
ellipsoid after its configuration was relaxed.

\begin{figure*}[tbp]
  \centering
  \includegraphics[width=1.15\textwidth]{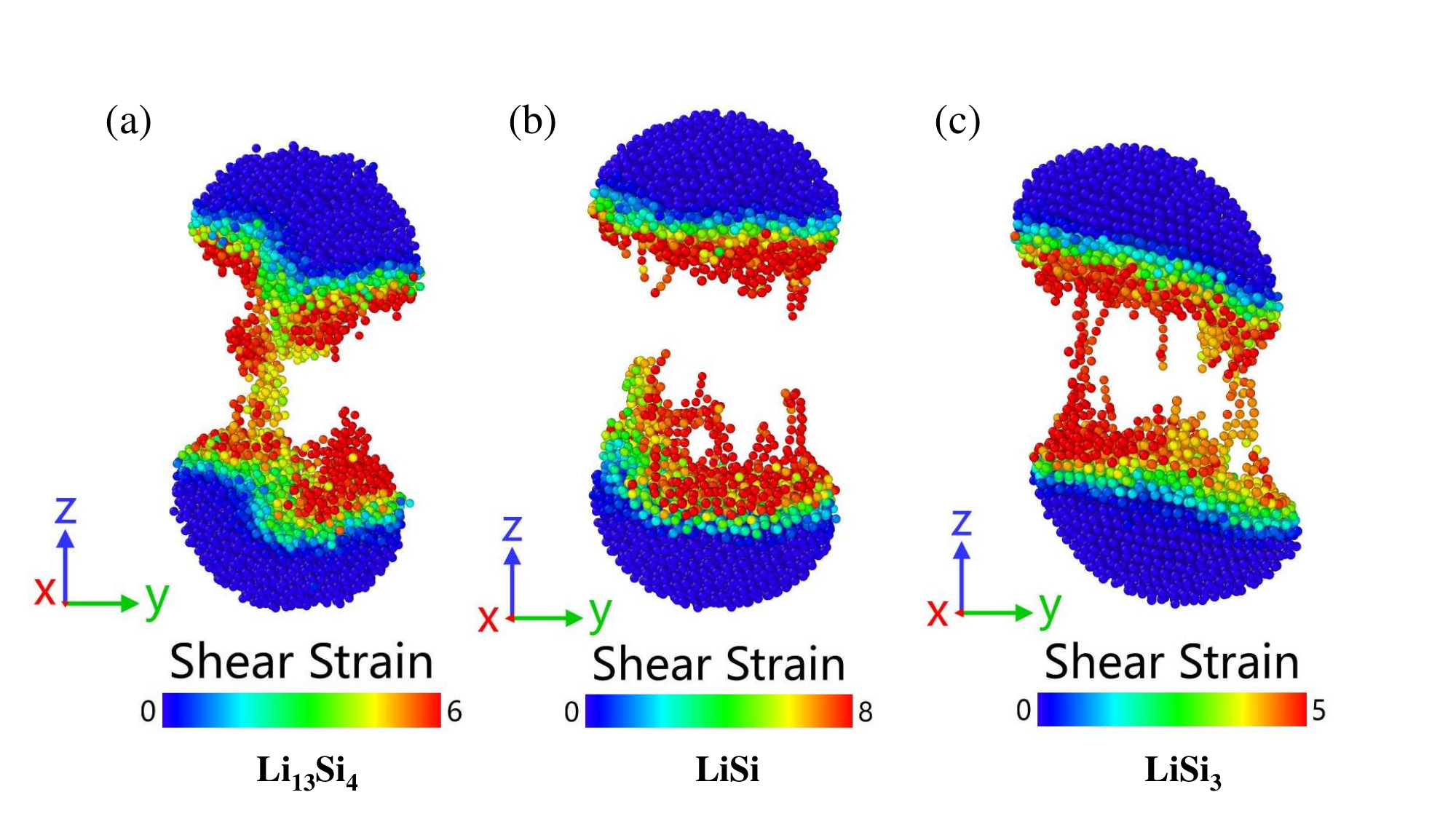}
  \vspace{-0.5\baselineskip}
  \caption{\label{fig: np}%
    \textbf{Final shear strain states of different nanoparticles after the stretching process.}
    \textbf{a}, \ce{a-Li_13Si_4}.
    \textbf{b}, \ce{a-LiSi}.
    \textbf{c}, \ce{a-LiSi_3}.
    }
\end{figure*}

From \textbf{Fig.~\ref{fig: nanoparticle}d} to \textbf{Fig.~\ref{fig: nanoparticle}f}, bond-breaking or fracturing behaviours are observed, with the bonds around centered atoms breaking first and gradually spreading to the outer shell or surface.
This is also observed in the bulk structure when the internal hole grows as described above.
The final states of different nanoparticles are shown in \textbf{Fig.~\ref{fig: np}}.
In the order of left to right, they are corresponding to the c, e, and h in \textbf{Fig.~\ref{fig: np-stress}}, respectively.
As shown in the figure, arrays of lines of atoms are observed around the round edge, meaning that they are the last ones to break in the stretching process and thus explain the reason for large plastic strains.


\subsection{Deformation mechanism and mechanical properties of stacked nanoparticles}

\begin{figure*}[tbp]
  \centering
  \includegraphics[width=1.15\textwidth]{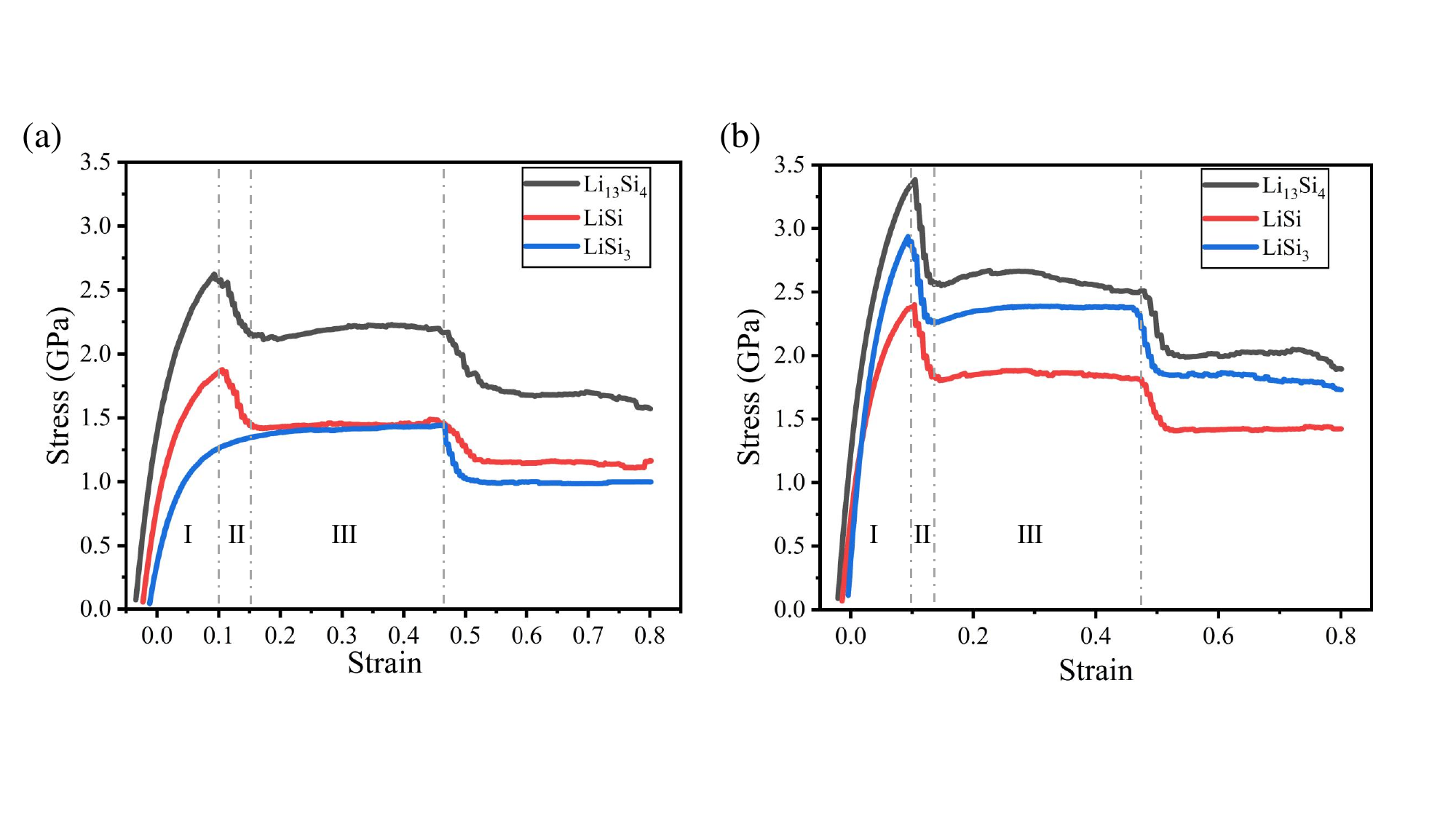}
  \vspace{-0.5\baselineskip}
  \caption{\label{fig: stack-stress}%
    \textbf{Stress-strain curve of stacked nanoparticles with different stacking numbers and different Li-Si compositions.}
    \textbf{a}, Stacking 50 nanoparticles.
    \textbf{b}, Stacking 80 nanoparticles.
    }
\end{figure*}

The stress-strain curve of stretching stacked nanoparticles with different compositions is shown in \textbf{Fig.~\ref{fig: stack-stress}}.
As can be seen from the figure, despite being stacked from different numbers of nanoparticles with different compositions, all structures' stress-strain curves show similar trends and exhibit typical elastic-plastic behaviours that could be divided into three stages, with the 50 stacked \ce{a-LiSi_3} nanoparticle as only exception.

In both stacking situations, stacked \ce{a-Li_13Si_4} nanoparticles have the maximum stiffness and highest tensile strength.
It is worth noting that for stacking 50 nanoparticles, the stiffness and ultimate tensile strength of \ce{a-LiSi} are larger than those in \ce{a-LiSi_3}, whereas for stacking 80 nanoparticles the situation is completely opposite.
Together, this means the mechanical properties are influenced by both the staking numbers and the Li-Si composition.

\begin{figure*}[tbp]
  \centering
  \includegraphics[width=1.15\textwidth]{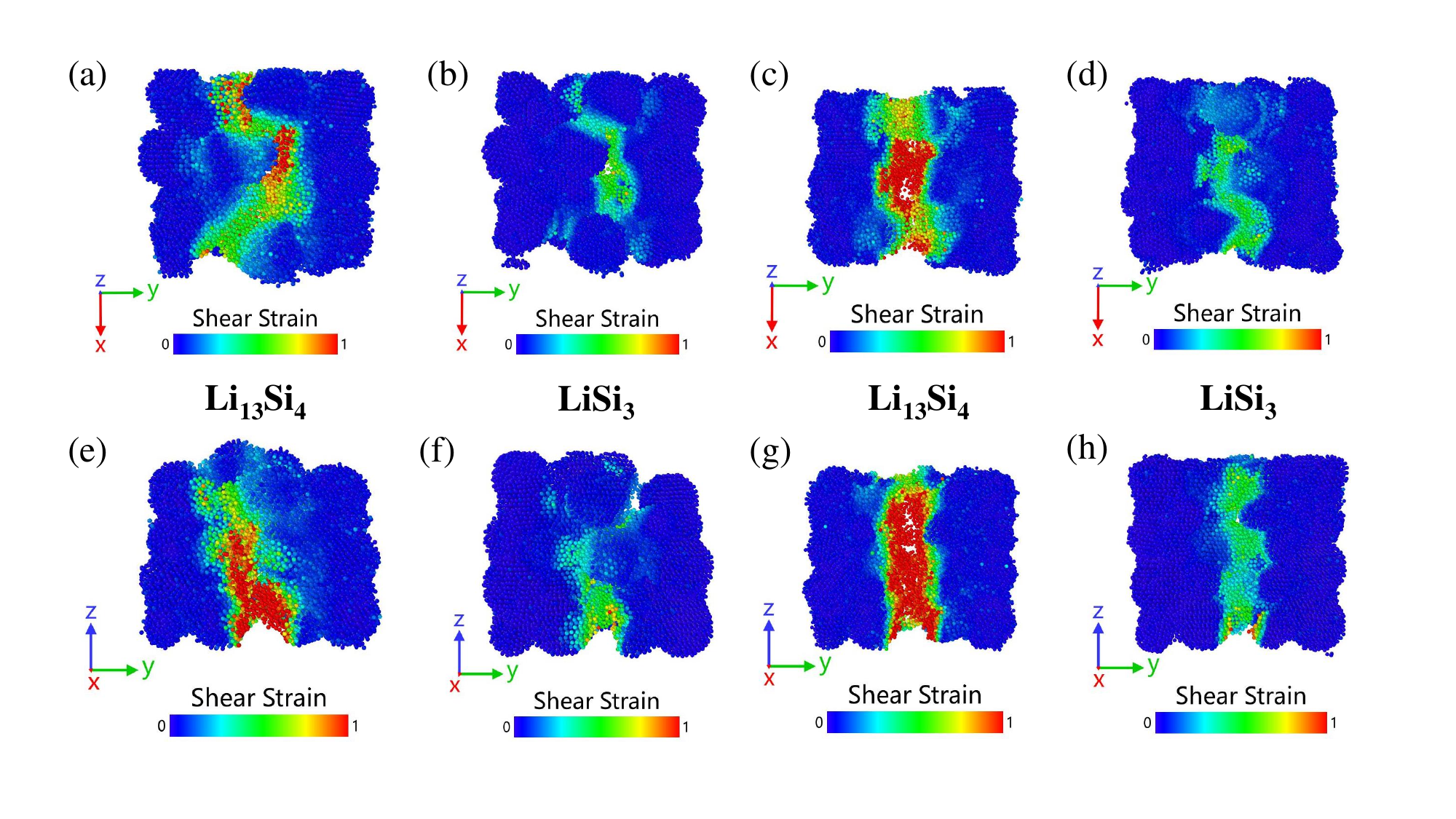}
  \vspace{-0.5\baselineskip}
  \caption{\label{fig: stack1}%
    \textbf{Morphology of different stacked structures at the end of stage \uppercase\expandafter{\romannumeral2} in the stretching process.}
    \textbf{a}, \textbf{b}, Distribution of shear strains in \ce{a-Li_13Si_4} and \ce{a-LiSi_3} structures stacked with 50 nanoparticles.
    \textbf{c}, \textbf{d}, Distribution of shear strains in \ce{a-Li_13Si_4} and \ce{a-LiSi_3} structures stacked with 80 nanoparticles.
    \textbf{e}, \textbf{f}, Another view for \textbf{a} and \textbf{b}.
    \textbf{g}, \textbf{h}, Another view for \textbf{c} and \textbf{d}.
    }
\end{figure*}

To relate the deformation mechanism to the stress-strain curve, the morphology of the deformations in \ce{a-Li_13Si_4} and \ce{a-LiSi_3} structures is shown in \textbf{Fig.~\ref{fig: stack1}}.
Since \ce{a-Li_13Si_4} and \ce{a-LiSi} structures deform in a similar way and the stress-strain curve of \ce{a-LiSi_3} structure is different than the others, \ce{a-Li_13Si_4} and \ce{a-LiSi_3} structures are chosen to compare the morphology.

First, as seen in \textbf{Fig.~\ref{fig: stack1}a} and \textbf{Fig.~\ref{fig: stack1}e}, the curved shear band could be clearly visualized which indicates that the cracking surface is not flat but curved.
The same phenomenon is also found in \textbf{Fig.~\ref{fig: stack1}b} and \textbf{Fig.~\ref{fig: stack1}f} for the \ce{a-LiSi_3} structure.
The biggest difference between \textbf{Fig.~\ref{fig: stack1}a}, \textbf{Fig.~\ref{fig: stack1}e} and \textbf{Fig.~\ref{fig: stack1}b}, \textbf{Fig.~\ref{fig: stack1}f} is the size of the shear band and thus the content of penetration.
The shear band in \ce{a-Li_13Si_4} structure is larger and penetrates through the whole cross-section, whereas in \ce{a-LiSi_3} structure the cross-section is not fully penetrated by the shear band.
This could explain why the stress-strain curve of \ce{a-LiSi_3} structure does not show the pattern of dropping as the other two structures in stage \uppercase\expandafter{\romannumeral2} in \textbf{Fig.~\ref{fig: stack-stress}a}.
The bonds around the centered atoms in the cross-section break more slowly than the rest of the systems.
Then, for stacking 80 particles, even though the shear bands appear to be flatter in \textbf{Fig.~\ref{fig: stack1}c} and \textbf{Fig.~\ref{fig: stack1}d} than in \textbf{Fig.~\ref{fig: stack1}g} and \textbf{Fig.~\ref{fig: stack1}h}, there are still some regions with curved shapes.
Notably speaking, for \ce{a-LiSi_3} structure, the shear band penetrates through the cross-section quickly so that the dropping would show up in this case.

\begin{figure*}[tbp]
  \centering
  \includegraphics[width=1.15\textwidth]{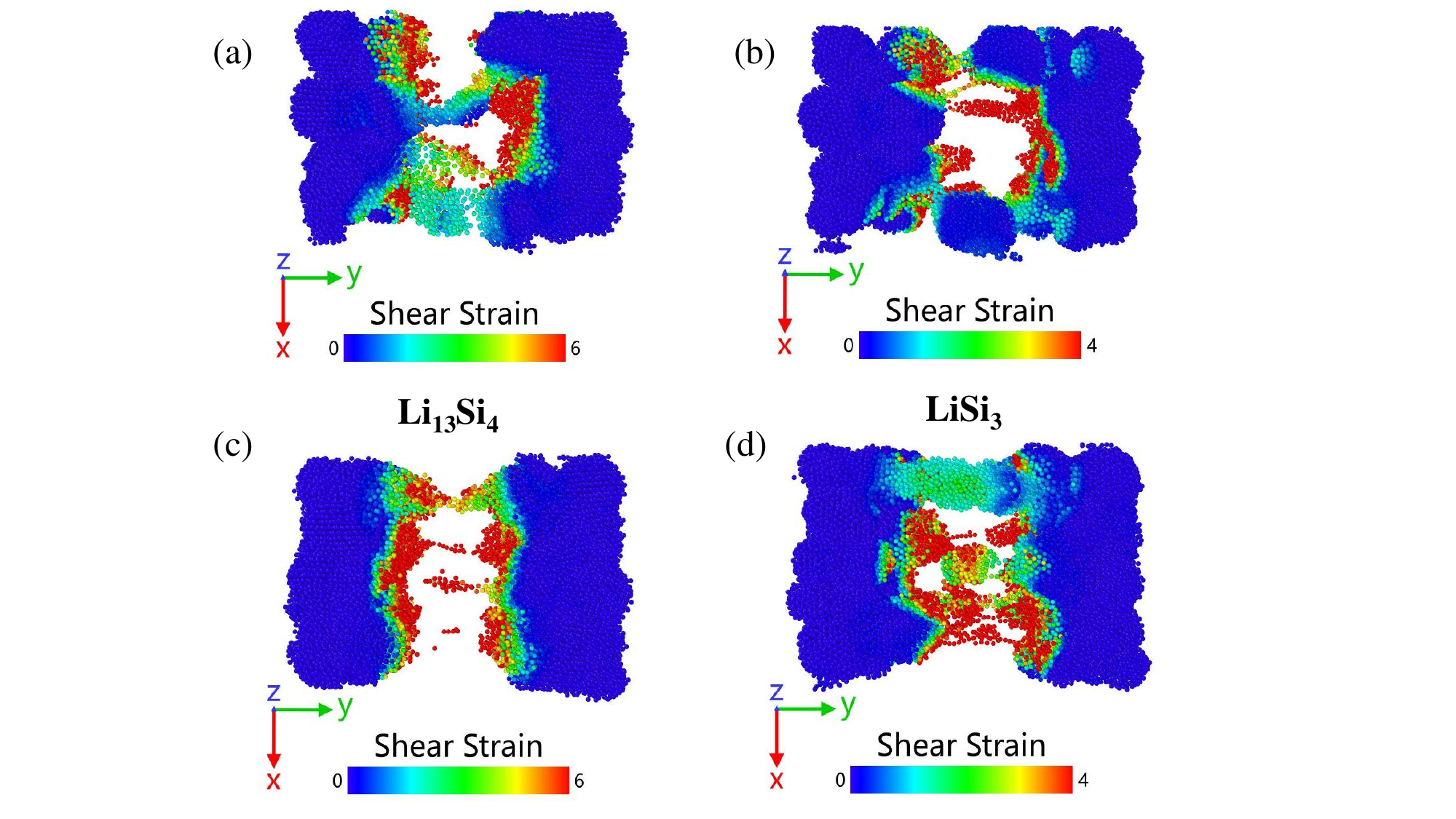}
  \vspace{-0.5\baselineskip}
  \caption{\label{fig: stack2}%
    \textbf{Final shear strain states of different structures at the end of stage \uppercase\expandafter{\romannumeral3} in the stretching process.}
    \textbf{a}, \textbf{b}, Final distribution of shear strains in \ce{a-Li_13Si_4} and \ce{a-LiSi_3} structures stacked with 50 nanoparticles.
    \textbf{c}, \textbf{d}, Final distribution of shear strains in \ce{a-Li_13Si_4} and \ce{a-LiSi_3} structures stacked with 80 nanoparticles.
    }
\end{figure*}

The final strain states of different stacked nanoparticles are shown in \textbf{Fig.~\ref{fig: stack2}}.
It can be seen that most fracturing behaviour happens within nanoparticles, whereas in \textbf{Fig.~\ref{fig: stack2}b} there is a nanoparticle detaching from the structure instead of cracking internally.

\pagebreak

\section{Conclusions}

In this work, a general and robust MLP was constructed based on a comprehensive data set that covers a variety of different Li-Si structures.
Various complex and realistic Li-Si models were constructed first and then the uniaxial stretching process was simulated by MD simulations to elucidate the deformation mechanisms and investigate the mechanical properties.
Based on the above results and discussion, the following conclusions could be drawn:

(1) The \ce{a-Li_xSi} bulk structures, nanoparticles with different compositions, and different-numbered stacked nanoparticles show typical elastic-plastic behaviours.
Yet, only \ce{a-Li_13Si_4} nanoparticle displays the obvious plastic yielding stage in the stress-strain curve.

(2) The amorphous Li-Si alloys show good ductility and toughness in all forms.
In the form of nanoparticles, the stiffness of \ce{a-Li_13Si_4} is largest, followed by \ce{a-LiSi_3} and \ce{a-LiSi} in the order of values.
Whereas for the ultimate tensile strength, the order would be as \ce{a-LiSi_3}, \ce{a-Li_13Si_4}, and \ce{a-LiSi}.
The stacking numbers of nanoparticles together with Li-Si compositions affect the mechanical properties of the Li-Si system since the order of stiffness and ultimate tensile strength would change for different structures.

(3) The deformation mechanisms in bulk and nanoparticle structures are visualized and elucidated on the atomic scale.
The plastic behaviours stem from the formation of shear bands whose evolution is elaborated for the bulk structure.
The fracturing pattern of the Li-Si system is clarified.
The origin of the fracturing is in the middle part of every structure due to the bond breaking.
The propagation of the fracturing is from inside to the outside until the cross-section is fully penetrated.

This work shows that MD simulations with MLPs can be used as tools to investigate the deformation mechanism and the origin of mechanical instabilities caused by fracturing on an atomic level.
The first principle-based MLPs are sufficiently accurate and efficient to a) model the large structure necessary to describe the deformation process and b) describe the breaking of bonds occurring in the course of the process.

By providing detailed insights into the different steps required to simulate the mechanical behaviour of Li–Si systems, this work aims to establish MLP-based simulations as a tool to understand the mechanical behavior in different materials systems at the atomic scale.



\section{Acknowledgements}
\vspace*{-0.5\baselineskip}
Zixiong Wei greatly acknowledges the financial support provided by the Chinese Scholarship Council. The project is part of the Dutch sector plans for scientific research and university education. We thank Jon Lopez-Zorrilla, Theophile Tchakoua, and Frank de Groot for their helpful discussions. Nong Artrith thanks the Dutch National e-Infrastructure and the SURF Cooperative for providing computational resources used in the implementation, benchmarking of the ænet-PyTorch package, and training MLPs.

\section{Notes}
\vspace*{-0.5\baselineskip}

The authors declare no competing financial interest.
This work made use of the free and open-source atomic energy network (ænet), ænet-PyTorch package. The source code can be obtained either from the ænet Web site (http://ann.atomistic.net) or from GitHub (https://github.com/atomisticnet/aenet-PyTorch). The MD simulations with MLPs input and output files can also be obtained from the GitHub (https://github.com/atomisticnet/XXX). The reference \ce{Li_xSi} data set can be obtained from the Materials Cloud repository
\\ (https://doi.org/10.24435/materialscloud:dx-ct). The data set contains atomic structures and interatomic forces in the XCrySDen structure format (XSF), and total energies are included as additional meta information.


%


\newpage
\bibliographystyle{aipnum4-1}

%


\appendix
\renewcommand{\thefigure}{S\arabic{figure}}
\renewcommand{\thetable}{S\arabic{table}}
\setcounter{figure}{0}
\setcounter{table}{0}


\end{document}